\documentclass[aps,prd,showpacs,nobibnotes,nofootinbib]{revtex4}
\usepackage{graphicx}
\usepackage{amsmath,amssymb,amsfonts, txfonts,stmaryrd}
\usepackage{epstopdf, setspace}
\usepackage{epsfig}
\include{psfig}
\input{epsf}
\def\ds{\displaystyle}
\def\sech{\, \textrm{\rm sech}\, }

\newtheorem{theorem}{Theorem}
\newtheorem{lemma}[theorem]{Lemma}

\begin{document}
\title{SO(10) domain-wall brane models}
\author{Jayne E. Thompson}
\email{j.thompson@pgrad.unimelb.edu.au}
\affiliation{School of Physics, The University of Melbourne, Victoria 3010,
Australia}
\author{Raymond R. Volkas}
\email{raymondv@unimelb.edu.au}
\affiliation{School of Physics, The University of Melbourne, Victoria 3010,
Australia}
\date{\today}
\begin{abstract}
We construct domain-wall brane models based on the grand-unification group SO(10), generalising the SU(5)
model of Davies, George and Volkas.  Motivated by the Dvali-Shifman proposal for the dynamical
localisation of gauge bosons, the SO(10) symmetry is spontaneously broken inside the wall.  We present
two scenarios: in the first, the unbroken subgroup inside the wall is ${\rm SU(5)} \times {\rm U(1)}_X$, and in the
second it is the left-right symmetry group SU(3)$\times$SU(2)$_L\times$SU(2)$_R\times$U(1)$_{B-L}$.  In both cases
we demonstrate that the phenomenologically-correct fermion zero modes can be localised to the wall,
and we briefly discuss how the symmetry-breaking dynamics may be extended to induce breaking to
the standard model group with subsequent electroweak breaking.  Dynamically localised gravity is
realised through the type 2 Randall-Sundrum mechanism.
\end{abstract}
\maketitle

\newpage
\section{Introduction}

Over the last decade or so there has been a great revival of
interest in models featuring extra dimensions of space. Kaluza-Klein
theory, originally proposed early in the 20th century, is the
forerunner of modern universal extra-dimension models
\cite{Kaluza:1921tu, Klein:1926tv}. These theories
hypothesise additional small extra dimensions of compact topology,
with all degrees of freedom unrestricted in their propagation.  An
alternative scenario sees some fields localised to a submanifold or
brane, while other fields propagate throughout the bulk.  In
particular, Arkani-Hamed, Dimopoulos and Dvali pointed out that
compact extra dimensions may be rather large in characteristic scale
if only gravity is able to propagate in the bulk
\cite{ArkaniHamed:1998rs}.

At about the same time, two seminal papers by Randall and Sundrum,
referred to hereinafter as RS1 and RS2, investigated brane-world
models with non-trivial spacetime curvature in the bulk (warped
metrics). RS1 is a model incorporating one compact extra dimension
with a topological identification $[y] \in S^1/Z_2$ and a
warped-metric geometry \cite{Randall:1999ee}. Two branes, denoted
Planck and TeV, are placed at the orbifold fixed points $\left[y\right] = 0$
and $[y]=r_c\pi$, respectively. The original setup had all
standard-model fields confined to the TeV brane, while the graviton
was dynamically-localised to the Planck brane. The metric warp
factor aggregates normalisation conditions so that the physical mass
scale, on the TeV brane, is related by an exponential hierarchy to
the free parameters in the Lagrangian. Thus it is possible to have
the Lagrangian parameters at a high scale while maintaining
comparatively small standard model masses, thereby solving the gauge
hierarchy problem. RS2 is obtained, effectively, by taking the TeV
brane infinitely far from the Planck brane and reflecting the
resulting spacetime about the Planck brane \cite{Randall:1999vf}.
The result is a model with an infinite extra dimension but still
with effective four-dimensional gravity dynamically induced on the
brane.  The standard model fields must now be taken to live on the
Planck brane. [See references \cite{Antoniadis:1990ew},\cite{Antoniadis:1998ig},\cite{Visser:1985qm} for other foundational work on extensions to the standard model using extra dimensions.]

RS2 is of considerable interest because the extra dimension is
treated analogously to the usual three dimensions of space, unlike
the compact extra dimension paradigm. However, pure RS2 leaves open
the question of what the brane is (D-brane of string theory?) and
why non-gravitational fields are localised to it.  Also, the extra
dimension is not exactly on the same footing as the other
dimensions, because the placement of the fundamental brane
explicitly breaks translational invariance along the extra
dimension.

Independently of the above, it had been proposed that our universe might be a domain wall
or other topological soliton \cite{Rubakov:1983bb}. See also \cite{Akama:1982jy},\cite{Gibbons:1986wg}.  In the codimension 1 case of a domain-wall, the setup
is quite similar to RS2.
Such a universe is brane-like, but it has finite thickness
and the origin of the brane is specified.  Brane formation is now an instance of
the spontaneous rather than the explicit breakdown of translational invariance.  Finally,
one can envisage that \emph{all} fields are localised for dynamical reasons to the brane,
not just the graviton as in pure RS2.  What, exactly, would such a theory look like?

Recently, one of the authors (RRV), proposed, together with Davies
and George, a specific theory (to be called the DGV model
hereinafter) that may realise the dynamical localisation of an
${\rm SU}(3)_C \times {\rm SU}(2)_W \times {\rm U}(1)_Y$ gauge theory plus
gravity to a solitonic domain-wall brane with one warped extra
dimension \cite{Davies:2007xr}. The DGV paper speculates that
Dvali-Shifman gauge field localisation (to be reviewed in a later
section) is implemented by confining ${\rm SU}(5)$ bulk gauge
dynamics. The purpose of this paper is to extend this model by
enlarging the bulk gauge group to ${\rm SO}(10)$.  This allows the
domain-wall localised gauge theory to be either ${\rm SU}(5) \times
{\rm U}(1)_X$ or the left-right symmetric model, depending on the
choice of Higgs potential.

In the next section, we discuss the qualitative features of this kind of domain-wall brane model,
as a warm up for the detailed constructions presented in sections \ref{sec:SO(10)breakstoSU(5)model} and \ref{sec:SO(!0)breakstoSU(3)timesSU(2)LtimeSU(2)RtimesU(1)}.  Section \ref{sec:Dvali-Shifman gauge field localisation} critically
reviews the Dvali-Shifman mechanism, while our final two sections contain a discussion of salient points for our model and conclusion.

\section{The Dvali-Shifman domain-wall brane setup}

The most basic structure in domain-wall brane models is a kink configuration for a scalar field.
The prototype kink is supplied by a ${\Bbb Z}_2$ invariant $\phi^4$ potential for a real scalar
field $\phi$. The Euler-Lagrange equations possess a solitonic kink
solution,
\begin{equation}
\phi(x^{\mu},y) = v \tanh(my),
\end{equation}
where $v$ is the ${\Bbb Z}_2$-breaking vacuum expectation value, while
$m$ is the inverse width of the kink. Both parameters are given
through the Higgs potential.  The extra dimension is described by
the coordinate $y$. The energy density profile for this solution is
localised around the zero of the kink ($y=0$). We say the scalar
field has condensed to form a domain wall. The region away from the
3+1-dimensional domain wall is called the bulk.  The configuration
is topologically stable because it asymptotes to vacuum expectation
values that spontaneously break a discrete symmetry, and are thus
disconnected from each other in the vacuum manifold.

The prototype domain wall must be embedded in a considerably richer
theoretical structure in order to produce a model that might have
realistic phenomenology.  In particular, the requirement that
massless gauge bosons be dynamically localised to the wall is most
plausibly met by invoking the Dvali-Shifman mechanism
\cite{Dvali:1996xe}.  This proposal will be critically reviewed in
the next section. It requires the unbroken gauge group, $H$, inside
the wall to be a subgroup of that in the bulk $G$.  The gauge bosons
of $H$ are then thought to be localised, provided that the bulk
gauge theory is in a confinement phase.

The DGV model uses $G = {\rm SU}(5)$ and $H = {\rm
SU}(3)_C\times {\rm  SU}(2)_W  \times {\rm U}(1)_Y$, thus potentially realising a
domain-wall localised standard model.  It is also a novel
reinterpretation of SU(5) grand unification. As with standard SU(5)
theories, the extension to SO(10) immediately suggests itself
because all the standard model fermions of a given family may be assembled
into a single irreducible representation, the $16$.  In this paper,
we thus start with $G = {\rm SO}(10)$.  We also need a discrete
symmetry outside of SO(10) for topological-stability reasons; we
choose ${\Bbb Z}_2$ for simplicity and economy.\footnote{An earlier attempt at ${\rm SO}(10)$ domain-wall brane models can be found in \cite{Shin:2003xy}, based on the clash of symmetries idea \cite{Davidson:2002eu}. However, these schemes are unable to produce phenomenologically acceptable fermion localisation \cite{Curtin:2006sv}.}

We therefore consider an ${\rm SO}(10)\times {\Bbb Z}_2$ gauge-invariant Lagrangian, eventually
to be coupled to a Randall-Sundrum warped 4+1-dimensional metric in order to dynamically localise the graviton.
Group-theoretically extrapolating the original argument
presented by Dvali and Shifman, an SO(10) singlet scalar field in conjunction with
an SO(10) adjoint Higgs is used to dynamically generate the required domain wall.  The singlet Higgs
takes on a kink configuration, while
the adjoint Higgs configuration is nonzero inside the wall, spontaneously breaking SO(10), and asympototes
to zero at $\lvert y \lvert = \infty$ thus restoring SO(10) in the bulk.
For quartic potentials, we shall show that the unbroken group inside the wall is ${\rm SU}(5) \times {\rm U}(1)_X$, while
the extension to sixth order permits the further breaking to the left-right group
${\rm SU}(3)_C \times{\rm SU}(2)_L\times{\rm SU}(2)_R\times{\rm U}(1)_{B-L}$.  If the Dvali-Shifman mechanism is operative,
then the gauge bosons of these subgroups are dynamically localised.

Chiral 3+1-dimensional fermions are localised by their interactions
with the background Higgs field configurations
\cite{Rubakov:1983bb}. Each fermion is confined around a
3+1-dimensional hypersurface parameterised by certain values for the
bulk coordinate $y$. These values differ according to fermion
species, because the adjoint Higgs configuration ``splits'' their
localisation points.

For completeness we discuss subsequently breaking ${\rm SU}(5) \times
{\rm U}(1)_X$ and, respectively, ${\rm SU}(3)_C \times {\rm SU}(2)_{L}
\times {\rm SU}(2)_{R} \times {\rm U}(1)_{B-L}$
down to the standard model on the domain wall brane. This will require additional
Higgs fields which must condense inside the domain wall and some fine tuning conditions to induce the required gauge hierarchies.

An important advantage of this type of 4+1-dimensional grand-unified theory is that the usual tree level mass relations between the quark and lepton masses no
longer appear. This is precisely because the fermions are split along the
extra dimension. Each fermion's 3+1-dimensional mass scale depends
on overlap integrals of the fermion's bulk profile with the extra-dimensional profile functions of the additional Higgs fields which we introduced to break the symmetry down to the standard model gauge group.  For the ${\rm SU}(5) \times {\rm U}(1)_X$ case, the bulk profiles for the fermions depend on their distinct ${\rm U}(1)_X$ charges, thus these overlap integrals contribute factors to the 3+1-dimensional mass parameters which are different for fermions in different ${\rm SU}(5) \times {\rm U}(1)_X$ representations. Hence the tree level mass relations are reduced to the ${\rm SU}(5) \times {\rm U}(1)_X$ subset of the normal ${\rm SO}(10)$ mass relations. A similar effect occurs for the left-right symmetric alternative model.

Before discussing the technical details of the above models, we need to pause and
talk briefly about using the Dvali-Shifman mechanism to confine
gauge fields to a codimension 1 domain-wall brane in 4+1-dimensions. We shall use this
opportunity to offer a brief heuristic review of the mechanism.

\section{Dvali-Shifman gauge field localisation}
\label{sec:Dvali-Shifman gauge field localisation}

It is straightforward to dynamically localise both chiral fermions and other scalar fields
to a domain-wall background.  But the localisation of massless gauge fields is a much
greater challenge.  The simple analogue of the fermion and scalar localisation mechanisms
can produce localised \emph{massive} spin-1 fields, but this is not what we require, both because
photons and gluons are massless, and because gauge coupling constant universality is lost
due to differing overlap integrals involving the extra-dimensional profile functions.

A very different mechanism is required, and the most well-known proposal is conceptually different from
the purely classical physics one may use to trap fermions and scalars.
 Dvali and Shifman consider a toy model based on an ${\rm SU}(2)$ Yang-Mills theory in
3+1-dimensions breaking to ${\rm U}(1)$ on a 2+1-dimensional domain
wall. The bulk is presumed to be in confinement phase with a
strong coupling regime in the infrared. This creates a mass gap of
order the bulk confinement scale $\Lambda_{{\rm conf}[{\rm SU}(2)]}$ in the vector boson spectrum, so
away from the wall the gauge fields form bound states with mass
$> \Lambda_{{\rm conf}[{\rm SU}(2)]}$.

On the domain wall the photon is free, however in order to
propagate in the direction transverse to the wall it must become
conglomerated into a massive ${\rm SU}(2)$ glueball. The energy cost
associated with moving off the wall is untenable for the massless
${\rm U}(1)$ gauge boson which effectively becomes trapped on the
2+1-dimensional topological defect.

In the 't Hooft-Mandelstam \cite{'tHooft:1981ht, Mandelstam:1974pi} dual superconductivity picture
\cite{ArkaniHamed:1998ww}, electric field lines from test charges on
the brane spread out like normal Coulomb fields along the brane
directions. However, when they meet the bulk they get repelled by
the dual analogue of the Meissner effect. Thus any two charges
located on the brane will interact via a 2+1-dimensional Coulomb
force at distances much greater than the thickness of the wall.

If the test charge is located in the bulk then the field lines are
no longer at liberty to spread out. Instead they form a flux
string which tunnels through to the brane and expels the field
onto the domain wall. This is the dual superconductor effect
corresponding to Abrikosov vortex formation. Hence, regardless of
the bulk localisation profiles of the fermions, the effective field
on the domain wall will exhibit the same 2+1-dimensional Coulomb
distribution.

The Dvali-Shifman argument can easily be extended to deal with a
more general gauge group $G$ breaking to some subgroup $H$ on the
domain wall. Under these circumstances it is assumed that the
confinement scale on the brane, $\Lambda_{{\rm conf}[H]}$, is
significantly lower than the bulk scale, $\Lambda_{{\rm conf}[G]}$.

However, extending the model to cope with a larger number of
dimensions is not easy. The problem is that non-Abelian gauge
theories are not renormalisable in more than 3+1-dimensions.

3+1-dimensional non-perturbative lattice simulations provide
qualified support for the hypothesis that the Dvali-Shifman
mechanism can trap gauge fields on a codimension-1 brane
\cite{Laine:2004ji}. But to our knowledge no one has tackled a
4+1-dimensional simulation. At this stage the Dvali-Shifman
mechanism remains an intriguing conjecture in 4+1-dimensions.

A necessary condition for the Dvali-Shifman mechanism is confinement
in the bulk. It is encouraging that lattice gauge theory simulations
for ${\rm SU}(2)$ pure Yang-Mills theory in 4+1-dimensions
demonstrate a first order phase transition at finite lattice spacing
as a function of the gauge coupling constant: for coupling strengths
above a critical value, the theory appears to be confining
\cite{Creutz:1979dw}. This result has been extended to pure ${\rm SU}(5)$ Yang Mills gauge theories in \cite{DPGeorge2009}.

However no conclusion can be drawn
about the continuum limit since our non-renormalisable theory has
problems with point-like interactions at high energies.
But this may not actually be a problem for our application.  All field-theoretic
brane models are necessarily low-energy effective theories due to their non-renormalisability,
so they are implicitly defined with an ultraviolet cutoff.  The finite lattice
spacing in the simulations is also an ultraviolet cutoff, so it appears
sensible to use the lattice gauge results to conclude that
confinement in 4+1-dimensional effective gauge theories can exist.

While this argumentation provides encouragement to pursue models based
on the Dvali-Shifman idea, it does not rigorously establish that any specific model works.
For one thing, no such model is a pure Yang-Mills theory.  Secondly, for each candidate theory one
would need to compute via lattice simulations the critical gauge coupling constant as a function
of the cutoff.  Since the critical coupling constant in 4+1 dimensions is itself of nonzero mass dimension,
one would then need to check that this value is compatible with other scales in the problem, such as the
inverse width of the domain wall and the effective grand unified symmetry breaking scale
inside the wall.  In particular, we need the SO(10) breaking scale to be higher than the
bulk confinement scale.

We find it interesting that the assumed validity of the Dvali-Shifman mechanism allows
plausible candidate domain-wall localised gauge theories to be readily constructed.  We believe
this provides good motivation to pursue more in-depth studies of the mechanism via lattice gauge theory.

\section{The ${\rm SO}(10) \rightarrow {\rm SU}(5) \times {\rm U}(1)_X$ model}
\label{sec:SO(10)breakstoSU(5)model}

Our goal is an ${\rm SU}(5) \times {\rm U}(1)_X$ gauge theory on the
3+1-brane with a full ${\rm SO}(10)$ unified theory in the bulk.
For simplicity we shall start with a Minkowski flat action. We then
compile our theory using the following algorithm:

\begin{itemize}
\item{construct a domain wall using a scalar ${\rm SO}(10)$ singlet;}
\item{use an adjoint Higgs to break ${\rm SO}(10)$ to ${\rm SU}(5)
\times {\rm U}(1)_X$ on the domain wall and invoke Dvali-Shifman
gauge field localisation;}
\item{confine zero mode chiral fermions;}
\item{add a Randall-Sundrum warped metric;}
\item{discuss breaking ${\rm SU}(5) \times {\rm U}(1)_X$ to the
standard model on the domain wall and electroweak symmetry
breaking.}
\end{itemize}

It is helpful to establish a consistent notation before we start.
The indices M and N run from 1 to 5, while $\mu, \vee$ denote coordinate indices belonging to the subspace 1 to 4. We distinguish the
coordinate $x^5=y$. Our convention for the metric signature is
${\rm diag}(+,-,-,-,-)$.~Elements of the set $\{\sigma_1, \sigma_2,
\sigma_3\}$ are used to label the $2 \times 2$ Pauli matrices,
or 3 dimensional Clifford algebra.

\subsection{Domain wall construction and gauge field localisation}
\label{sec-domainwallconstructionandgaugefieldlocalisation}

We consider a Higgs sector constituted by an
${\rm SO}(10)$ Higgs singlet $\phi \sim 1$ and an adjoint Higgs
${\cal X} \sim 45$. We impose the discrete $Z_2$ symmetry, $y \rightarrow -y$, $\phi \rightarrow - ~\phi$, ${\cal X} \rightarrow -{\cal X}$. The Higgs potential is
\begin{equation}
V_{\cal{X}, \phi}   =  - \frac{\mu^2}{2}
{\rm Tr}{\cal X}^2 + \frac{\lambda_1}{4}({\rm Tr}{\cal X}^2)^2
 +  \frac{\lambda_2}{4}{\rm Tr}{\cal X}^4 +
\frac{\kappa}{2}\phi^2{\rm Tr}{\cal X}^2 + \frac{\lambda}{4}(\phi^2
- \vee^2)^2,
\label{eq: fourthorderhigspotential}
\end{equation}
\noindent{where we have truncated our potential at fourth order. We argue that the general expression for $V_{\cal{X}, \phi}$ can be expanded as a polynomial in the fields ${\cal X}$ and $\phi$ where higher order terms are suppressed in an effective low energy theory by their dimensionful coupling constants. We include 4th order terms because orthogonality of the ${\rm SO}(10)$ generators remonstrates that truncating our potential at second order would have introduced an accidental symmetry whereby each of the 45 ${\rm SO}(10)$ adjoint-Higgs field components could be transformed independently under ${\rm SO}(10)$. Since our theory is non-renormalizable in 4+1-dimensions and is therefore considered to be effective only up to an ultraviolet cut off we do not expect pathologies from the dimensionful coupling constants. The adjoint Higgs multiplet is represented by the $10\times 10$
anitsymmetric matrix ${\cal X} = \sum_i T_i {\cal X}_i$ where the
$T_i$ are the generators of the fundamental representation normalized so the ${\rm Tr} \left(T_i T_j \right) = \frac{1}{2} \delta_{ij}$. In a certain regime of parameter space
the vacuum manifold is $({\cal X} = 0, \phi = \pm \vee)$. Solitonic
solutions to the Euler-Lagrange equations must obey Dirichlet
boundary conditions, where we require both Higgs fields to asymptotically
approach vacuum configurations.
We choose the boundary condition to be $(0, -\vee)$ at $y =
-\infty$ and $(0, \vee)$ at $y = + \infty$.}

Since the vacuum manifold is not connected, the relevant homotopy
group $\pi_0({\rm SO}(10) \times {\Bbb Z}_2/{\rm SO}(10))$ is non-trivial.
Our topological boundary conditions exploit this. Any solution to
the Euler-Lagrange equations satisfying these boundary conditions
cannot spontaneously evolve into a solution from another
topological sector, irrespective of relative energy densities. Id
est, any map interpolating between the two disconnected pieces cannot be deformed continuously to give the trivial map from $S^0$ to
${\rm SO}(10) \times {\Bbb Z}_2/{\rm SO}(10)$.

Under these conditions, the stable solution to the Euler-Lagrange
equations is the minimum energy density solution belonging to this
topological sector. Like Dvali and Shifman, we identify stable
solutions by checking the dynamical evolution of perturbative
linear modes. We also checked our results numerically to account
for higher order effects. The energy density of each solution is
dependent on the coupling constants, hence different solutions are
stable in different coupling constant regimes.

We find that $\lambda_2 = 0$ is a bifurcation point for the manifold
of solutions to the Euler-Lagrange equations. When $\lambda_2 > 0 $
the ${\rm SO}(10)$ adjoint field will condense about $y = 0$, so that
the nonzero components of ${\cal X}$ arrange for ${\rm SO}(10)
\rightarrow {\rm SU}(5) \times {\rm U}(1)_X$ on the domain wall. When
$\lambda_2 < 0$ our analysis indicates that to first order the
${\cal X}$ field components will try to adopt a configuration which
breaks ${\rm SO}(10) \rightarrow {\rm U}(1)^5$ on the domain wall. A more
detailed discussion is given in the Appendix (\ref{sec-apendix}).

Let us take a closer look at the $\lambda_2 > 0$ scenario. Solutions
to the Euler-Lagrange equations persist for a wide range of
parameter values.  Purely for the sake of convenience, however,
we choose to focus on a concrete analytic solution that exists
provided the parameters obey
\begin{equation}
(4\mu^2-2\kappa\vee^2 + \lambda\vee^2)(20\lambda_1 + 2\lambda_2)-5\kappa(4\mu^2-\kappa\vee^2)
=0.
\label{eq:conditionfforbackgroundsolution}
\end{equation}
We emphasise that this is not a fine-tuning condition.  Rather, it defines
a special slice through parameter space that happens to admit an analytical solution.

Under these circumstances the Euler-Lagrange equations have a
stable solution of the form,
\begin{equation}
\phi(y) = \vee\tanh(my), \qquad {\cal X}_{1}  = A{\rm sech}(my),
\label{eq-SU(5)solutiontoeulerlagrangeequations}
\end{equation}
\noindent{with $A^2 = \frac{40\mu^2 - 10\kappa\vee^2}{10\lambda_1 +
\lambda_2}$, $m^2 = -2\mu^2 + \kappa \vee^2$ and ${\cal X}_{1}$ is
associated with the ${\rm U}(1)_X$ generator which we embed inside
${\rm SO}(10)$ according to $T_1 = \frac{1}{\sqrt{20}}{\rm diag}(\sigma_2,\sigma_2,
\sigma_2,\sigma_2, \sigma_2)$, where $\sigma_2$ is the second
Pauli matrix. Aside from having a closed form, this solution is
convenient since the dynamical equations for the first order terms in a perturbative expansion about this solution can be transformed
into hypergeometric differential equations. We are able to solve
the equations exactly and show that all perturbative modes are
oscillatory. So, the solution is stable against further
condensation of the remaining $44$ ${\cal X}$ field
components.}

As promised, this solution spontaneously breaks the ${\rm SO}(10)$
gauge symmetry down to ${\rm SU}(5) \times {\rm U}(1)_X$ on the domain
wall. So if we assume the Dvali-Shifman mechanism to be effective,
that is assume the 4+1-dimensional bulk ${\rm SO}(10)$ gauge theory
is in a confining phase which is valid up to some ultraviolet cutoff
$\Lambda_{\rm UV}$, then ${\rm SU}(5) \times {\rm U}(1)_X$ gauge field
localisation will follow as a consequence of the background Higgs
field configuration.

Having established the existence of a stable Dvali-Shifman domain wall
solution, we now turn our attention to the localisation of fermions.

\subsection{Localising Fermions}
\label{sec:Localisingfermions}

We extend the Dirac algebra to encompass 4+1-dimensions by
identifying  $\Gamma^{M} = (\gamma^{\mu}, -i\gamma^5)$, where $\{ \Gamma^N, \Gamma^M \} = 2 \eta^{MN}$.
The fermions are contained in a $\Psi \sim 16$
representation of ${\rm SO}(10)$, and have the Lagrangian
\begin{equation}
{\cal L}_{{\rm Yukawa}} =  i\bar{\Psi}\Gamma^M\partial_M\Psi -
h_{\cal X}\bar{\Psi} \tau^a\tau^b {\cal X}_{ab}\Psi -
h_{\phi}\phi\bar{\Psi}\Psi,
\end{equation}
where ${\cal X}_{jk}$ is the $j,k$th entry of the $10 \times 10$ antisymmetric matrix used to represent ${\cal X}$ and $\tau^j$ is a member of the 10 dimensional Clifford algebra
\cite{Li:1973mq}. Specifically the Clifford algebra of dimension $n$
is constituted by any $n$ matrices satisfying $\{ \tau^j, \tau^k\} =
2\delta_{j,k} ~~ \forall j,k \in \{1, \dots n\}$.

We choose to adopt the explicit form
\begin{eqnarray}
\tau^{2k-1} & = & 1 \varotimes \dots \varotimes 1 \varotimes
\sigma_1 \varotimes  \sigma_3 \varotimes \dots
\varotimes \sigma_3,\\
\tau^{2k} & = &  \underbrace{1 \varotimes \dots \varotimes 1}_{k-1}
\varotimes ~\sigma_2 \varotimes  \underbrace{\sigma_3 \varotimes
\dots \varotimes \sigma_3}_{5-k},
\end{eqnarray}
\noindent{where it is understood that the index k runs over $\{1,
\dots 5\}$.}

To maintain the discrete reflection symmetry we now require $\Psi \rightarrow i \Gamma^5 \Psi$ in conjunction with $y \rightarrow -y$, $\phi \rightarrow - \phi$ and ${\cal X} \rightarrow -{\cal X}$.

Each spinor $\Psi_i(x,y)$, where the index $i$ denotes the different
irreducible ${\rm SU}(5) \times {\rm U}(1)_X$ components of the $16$, can be decomposed in terms of a complete set of simultaneous eigenfunctions $\psi_{n Li}(x)$ and $\psi_{n Ri}(x)$ for the $3+1$-dimensional Dirac Hamiltonian and (to remove the degeneracy) the 3+1-dimensional chirality operator $\gamma^5$. The basis states which appear with non zero coefficients in our expansion will be the 3+1-dimensional chiral zero mode, $\psi_{0L}$ or
$\psi_{0R}$ depending on the case, plus a finite number of discrete massive modes, as well as a continuum of massive modes, as per
\begin{equation}
\Psi_i(x,y) = \sum\hspace{-0.4 cm}\int_n \{f_{n Li}(y)\psi_{n Li}(x)+f_{n
Ri}(y)\psi_{n Ri}(x)\},
\label{eq-fermionspectrum}
\end{equation}
where the sum is understood to include an integration over the continuum.
This is a dimensional reduction or generalised Kaluza-Klein procedure
whereby the dependence on the $y$ coordinate of $\Psi_i (x,y)$ is subsumed into a complete set of mode functions or profiles $f_{n, L/R, i}$, and the
4+1-dimensional field is redescribed as an infinite tower of 3+1-dimensional fields.
The discrete mode functions are square-integrable, while the modes from the continuum are delta-function
normalisable.

\begin{figure}[htbp]
\begin{center}
\epsfbox{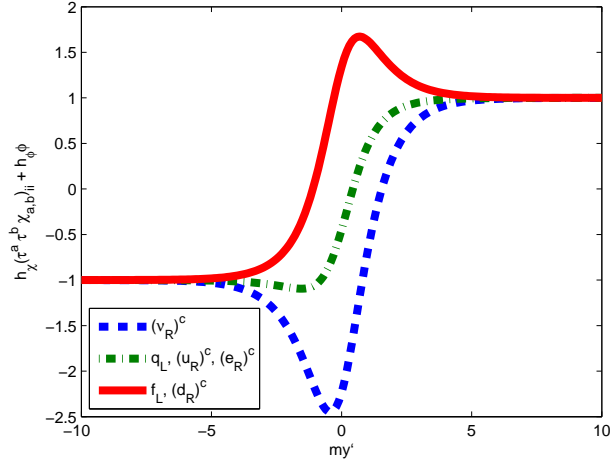}\\
 \caption{The graph displays the integrand in the exponent of equation (\ref{eq:fermionprofiles}) which gives the extra dimensional profiles of the left handed zero modes associated with first generation fermions in the ${\rm SU}(5) \times {\rm U}(1)_X$ brane world model. The graph shown here corresponds to the parameter choices $h_{\phi} = h_{\cal X} =1$ and the flat space background solution to the Euler-Lagrange equations ${\cal X}_1 = A{\rm sech}(my)$, $\phi = \vee{\rm tanh} (my)$ with $A = \vee = 1$. 3+1-dimensional left chiral fermions will be confined to the hyperplane corresponding to the zero of their integrand.}
\label{SU(5)bulkfermionprofiles}
\end{center}
\end{figure}

We want the 4+1-dimensional Dirac fermions in the $16$ of SO(10) to supply 16 left-handed 3+1-dimensional
zero modes to be identified with a family of quarks and leptons.  To display these modes, it suffices to
truncate the full mode decomposition to just the chiral zero mode term $f_{0Li}(y) \psi_{0 Li}(x)$.  This
ansatz is now substituted into 4+1-d Dirac equation with the background domain-wall configuration playing
the role of the mass term:
\begin{equation}
0  =  i\Gamma^M\partial_M  \Psi(x,y)
 -  h_{\cal X} \tau^a
\tau^b {\cal X}_{ab}(y) \Psi(x,y)
 -  h_{\phi}  \phi(y) \Psi(x,y).
\end{equation}
The separation of variables allows us to isolate the dependence on the
$y$-coordinate, after requiring that each of the $\psi_{0Li}$ be left chiral and obey the usual massless 3+1-dimensional
Dirac equation. Fortunately, for  ${\cal X} = A {\rm sech}(my)T_1$ the
matrix $\tau^a\tau^b {\cal X}_{ab}(y)$ is
diagonal. We label the diagonal entry belonging to row $i$ and column $i$ by $(\tau^a \tau^b {\cal
X}_{ab}(y))_{ii}$, where there is no intended sum over the index $i$. The differential equations decouple and the bulk
localisation profiles for the fermions are then easily found to be
\begin{equation}\label{eq:fermionprofiles}
f_{0Li}(y) = N_i e^{- \int^y_{y_0}{dy' \hspace{0.5mm} h_{\cal X}(\tau^a \tau^b {\cal
X}_{ab}(y'))_{ii} + h_{\phi} \phi(y')}},
\end{equation}
where $N_i$ is a normalisation constant.
We choose the signs of the coupling constants, $h_{\cal X}$ and
$h_{\phi}$, so that the integrand cuts the axis with a positive
slope. The asymptotic behaviour of the kink implements the
localisation of each fermion to a 3+1-dimensional hyperplane
coplanar with the zero of its respective intergrand, which occurs at $y = y_0$. Due to
coupling to the ${\rm SO}(10)$ adjoint Higgs, fermions belonging to
different ${\rm SU}(5)$ representations are localised around different
parallel hyperplanes in the bulk, see Figure \ref{SU(5)bulkfermionprofiles}.
\def\epsfsize#1#2{0.8#2}

These 3+1-dimensional localised massless fermions are our
${\rm SU}(5)$ brane world matter. In flat spacetime there is a mass
gap between the zero mode and the lowest allowed massive
3+1-dimensional mode. Hence, at low energies, interactions between
3+1-dimensional zero mass modes can only produce other zero mass
fermions \cite{Rubakov:2001kp}. This, in conjunction with the
Dvali-Shifman gauge boson localisation conjecture, produces a
candidate low-energy theory with dimensional reduction down to 3+1
dimensions.

We need to explain the physical interpretation we give to individual terms in our mode decomposition. Equation (\ref{eq-fermionspectrum}) is simply a convenient way of projecting the 4+1-d field onto a collection of 3+1-d coefficients $\psi_{n,L\setminus R,i}(x)$ by resolving the y-dependence of the field in terms of a complete set of orthonormal functions $f_{n,L\setminus R,i}(y)$ which span the 1-dimensional rigged Hilbert space. We have done this because it allows us to easily see how 3+1-dimensional states with definite mass feel the classical background. It is necessary to treat the different 3+1-dimensional eigenfunctions of $\gamma^5$ independently because we have associated $\Gamma^5$ with $\partial_y$. However mathematically equation (\ref{eq-fermionspectrum}) is simply a convenient way of writing the field.

Since the physics is basis independent, we can choose to expand a general field $\Xi(x,y) = \Sigma_n g_n(y) \xi_n(x)~(x \equiv x^{\mu})$ in terms of any complete set of states $\{g_n(y)\}$. For example: if the $\Xi(x,y)$ is an ${\rm SO}(10)$ gauge singlet then without loss of generality, we may conveniently choose the $\{g_n(y)\}$ so that the coefficients $\{\xi_n(x)\}$ are solutions to the massive 3+1-dimensional Klein-Gordon equation. In this case the profile functions $g_n(y)$ appearing in the above expansion correspond to physically states of the effective 3+1-dimensional theory which propagate as free particles in the confining bulk ${\rm SO}(10)$ gauge theory. In our case the $\psi_{n,L\setminus R,i}(x)$ are charged under ${\rm U}(1)_X$ and will be forced, by a confining bulk, to propagate as constituent particles of a bound ${\rm SO}(10)$ gauge singlet. This is conceptually akin to writing the QCD lagrangian in terms of quarks and gluons with the implicit understanding that the propagating states are hadrons. Since ${\rm SO}(10)$ is broken inside the wall, the 3+1-d fields behave in a manner consistent with the standard model gauge interactions.

We argue that because our zero mode profile functions are sharply peaked around the domain wall and the ${\rm SO}(10)$ confinement dynamics are suppressed here, our classical localisation profiles give us a reasonable first order approximation for the landscape of these low energy particles. See \cite{Davies:2007xr}, \cite{DPGeorge2009} for more details.

\subsection{Adding Warped Gravity}

The last vital component of our model is warped gravity. We search
for solutions to the Einstein-Klein-Gordon equations for the
action
\begin{equation}
S = \int {d^5 x \sqrt{G}\{-2M^3R - \Lambda + {\cal T} -  V_{{\cal X} \phi} +{\cal L}_{{\rm Yukawa}} \}},
\end{equation}
where $G$ is the determinant of the five-dimensional
metric tensor, $M$ is the 5-d fundamental gravitational mass
scale, R is the Ricci scalar, $\Lambda$ is the bulk cosmological
constant and ${\cal T}$ is simply the gauge field and Higgs boson
kinetic terms. We would like a solution with 3+1-dimensional
Minkowski space and zero mode gravitons localised on our field
theoretical brane. To achieve a Minkowski metric on the brane we impose a fine tuning condition on the bulk cosmological constant. The necessity of balancing the bulk cosmological constant against the brane tension is a feature of Randall-Sundrum like models. At the same time we need the
qualitative forms of both ${\cal X}$ and $\phi$ to be similar to the flat-space case.
Substituting a Randall-Sundrum warped metric ansatz,
\begin{equation}
ds^2 = e^{-p(y)/6M^3}\eta_{\mu\nu} dx^{\mu}dx^{\nu} - dy^2,
\label{eq-warpedmetricansatz}
\end{equation}
into the Einstein-Klein-Gordon equations produces four
coupled second order differential equations which exhibit appropriate
solutions for a wide range of parameter values. By choosing, as before purely for convenience, our
parameters to lie on the manifold,
\begin{eqnarray}
\begin{array}{r c l}
6m^2 M^3 + \vee^2m^2 - 3M^3\lambda\vee^2 + 3M^3 \kappa\vee^2& = &0,  \\
-3m^2M^3 - \vee^2m^2 - 6 M^3\mu^2 + 3 M^3\kappa\vee^2& = & 0, \\
20\lambda_1 + 2\lambda_2 - 10\kappa + 5 \lambda & = & 0, \\
\end{array}
\end{eqnarray}
\noindent{we find an analytic solution of the form:}
\begin{equation}
p(y) =  \vee^2 \ln({\rm cosh}[my]), \qquad
\phi(y)  =  \vee \tanh[my], \qquad
{\cal X}_1  =  2\vee{\rm sech}[my].
\label{eq-eienstienequationsolutions}
\end{equation}
This solution is consistent with all the localisation properties
mentioned above. It provides the same Higgs configuration which we
postulate induces Dvali-Shifman ${\rm SU}(5) \times {\rm U}(1)_X$
gauge field localisation on the brane.

In changing from flat space to a warped metric we must include vielbeins in the fermion action so that fields written at each point in terms of a local Lorentz coordinate system in the tangent space transform correctly under general coordinate transformations \cite{Weinberg:1972}. We have also included the spin connections to ensure the derivative is covariant under Lorentz transformations. With a metric-geometry described by (\ref{eq-warpedmetricansatz}) the fermionic terms in the action must be rewritten in terms of the vielbeins and spin connections
\begin{align}
V^{\mu}_A &= \delta^{\mu}_A e^{p(y)/12M^3}, & \omega_{\mu} &=  \frac{i}{24M^3}p'(y)e^{-p(y)/6M^3}\gamma_{\mu}\gamma^5,\\
V^5_A  &= \delta^5_A, & \omega_5  &= 0,
\end{align}
\noindent{where capital letters from the start of the Latin alphabet have been used to label the vielbeins' Lorentz indices and the Greek alphabet characters $\mu, \nu$ are coordinate indicies. Explicitly these become incorporated into the spin-covariant derivative $D_N = \partial_N +\omega_N$ and gamma matrices $\Gamma^N = V^N_A \Gamma^A$. With this simplified notation the fermions contribute to the warped space 5-d action according to:
\begin{equation}
S_{\psi} = \int d^5 x \sqrt{G}\{i\bar{\Psi} \Gamma^N D_N \Psi  -
h_{\cal X}\bar{\Psi} \tau^a\tau^b {\cal X}_{ab}\Psi -
h_{\phi}\phi\bar{\Psi}\Psi \}.
\label{eq-fivedimensionalfermionaction}
\end{equation}
We use an analysis presented in \cite{Davies:2007tq} to qualitatively demonstrate that if we decompose $\Psi(x,y)$ according to (\ref{eq-fermionspectrum}), then in warped space time there is no mass gap between the zero mode left handed fermion and the continuum of 3+1-dimensional massive left and right chiral modes. To make this statement transparent we use separation of variables in the Dirac equation, to obtain eigenvalue equations for the extra dimensional profiles. However we now consider the extra dimensional profiles of the massive modes, $f_{nLi}$, as well as the zero mode. These satisfy the equations:
\begin{eqnarray}
\begin{array}{r c l r}
0& = &6M^3 f_{0Li}'- p' f_{0Li} + 6M^3 g_j\phi_j f_{0Li}& \hspace{1 cm}  n=0  \\
24M^6 m_n^2 e^{p(y)/6M^3} f_{nLi}&=&-24M^6 f_{nLi}'' + 10M^3 p'f_{nLi}'
 +  [4M^3p'' -  {p'}^2   +  24M^6 W]f_{nLi}& \hspace{1 cm} n>0\\
\end{array}
\label{eq-warpedgeometrydiracequationextradimensionaldependence}
\end{eqnarray}
\noindent{where the generic scalar field coupling term $g_j\phi_j = h_{\cal X}(\tau^a \tau^b {\cal
X}_{ab})_{ii} + h_{\phi} \phi(y)$ has been used to simplify the expression for $W= (g_j\phi_j)^2- g_j\phi_j'+  g_j\phi_j\frac{p'}{12M^3}$ and $m_n$ is the mass of the nth 3+1-dimensional chiral mode. That is:}
\begin{equation}
i\gamma^{\mu}\partial_{\mu} \psi_{nLi} = m_n\psi_{nRi}, \qquad i\gamma^{\mu}\partial_{\mu}\psi_{nRi} = m_n\psi_{nLi}.
\end{equation}
We can solve for the zero mode extra dimensional profile immediately,
\begin{equation}
f_{0Li}(y)  =  N_0 e^{-\int^y_{y_0} dy'~ {g_j\phi_j - p'/6M^3 } }.
\label{eq-warpedgeommetryzeromode}
\end{equation}
For the solution presented in (\ref{eq-eienstienequationsolutions}) there is a normalised 3+1-dimensional left handed zero mode confined to the plane parameterized by $m y = {\rm ln} \left[ \textrm{root} \left(15 h_{\phi} M^3 t^2 + \left(6 \sqrt{5} h_{\cal X} \Sigma_{a,b}\left(\tau^a\tau^b\right)_{ii}\left(i\delta_{a,b+1}-i\delta_{a,b-1}\right) M^3 - 5 \vee m \right)t -15h_{\phi}M^3\right)\right]$ provided $6M^3 h_{\phi} - \vee m> 0$.

For an arbitrary point in the parameter space which will give rise to a different solution to the Einstein-Klein-Gordon equations we can still say something about the existence of a confined zero mode fermion. This is because square integrability of (\ref{eq-warpedgeommetryzeromode}) depends on the asymptotic properties of $p$ and $\phi_j$. We know that in the case of a scalar field Lagrangian, solving the Einstein-Klein-Gordon equations in a vacuum yields a warp factor $p \propto \lvert{y}\lvert$, hence we expect $p$ to approach $\lvert{y}\lvert$ as $y \rightarrow \infty$. We also impose the boundary conditions $lim_{y\rightarrow \pm \infty} {\cal X} = 0 $ and $lim_{y \rightarrow \pm \infty} \phi = \pm \vee$, thus any solution to the Einstein-Klein-Gordon equations will have the same asymptotic form as (\ref{eq-eienstienequationsolutions}).

For a continuous, bounded warp factor and Higgs field expressions with the same asymptotic form as (\ref{eq-eienstienequationsolutions}) there is a delta function normalisable zero mode, confined to the y-plane parameterized by the zero of the integrand in (\ref{eq-warpedgeommetryzeromode}), provided $6M^3 h_{\phi}-\vee m  > 0 $. So far our results are consistent with the flat space case. However we must now attend to the massive chiral 3+1-dimensional modes.

Previously we argued that the presence of a mass gap between the zero mode and the lowest discrete mode explained why electroweak scale experiments could not detect the tower and continuum modes. It has been shown, by \cite{Davies:2007tq}, that in Randall-Sundram warped space the continuum modes start from zero mass. To see this we make the substitution $\tilde{f}_{nLi} = e^{-p(y)/6M^3} f_{nLi}$ in (\ref{eq-warpedgeometrydiracequationextradimensionaldependence}) and change variables to conformal coordinates, $\frac{dz}{dy} = e^{p(y)/12M^3}$, to obtain
\begin{eqnarray}
\left[- \frac{d^2}{dz^2} + e^{-p(y(z))/6M^3} W \right]\tilde{f}_{nLi} & = & m_n^2 \tilde{f}_{nLi}.
\end{eqnarray}
The change to conformal coordinates, $f: y \rightarrow z$, is a diffeomorphism from ${\Bbb R}$ into a connected segment of the real line so we can analyze the potential $e^{-p(y(z))/6M^3} W$ as a function of $y$ and and interpret the results as corresponding to the z-coordinate space factored through the mapping $f^{-1} : z \rightarrow y$. This is well defined since $f^{-1}$ is bijective.

Using our Einstein-Klein-Gordon solutions (\ref{eq-eienstienequationsolutions}) we can easily see that $e^{-p(y)/6M^3} W $ asymptotes to zero. Thus we have a continuum of delta function normalized modes $\tilde{f}_{nLi}(z)$ starting at zero 3+1-dimensional mass, $m_n$, which approach plane waves asymptotically. This translates into a continuum of eigenfunctions, $f_{nLi}(y)$ $\forall m_n > 0$, which are delta function normalized with respect to the weight function $e^{-p(y)/4M^3}$. Davies and George \cite{Davies:2007tq} argue that these are precisely the normalization conditions required to reduce the kinetic term in the 4+1-dimensional action (\ref{eq-fivedimensionalfermionaction}) to its regular 3+1-dimensional counterpart. We take this as the condition for proper normalization of the profile functions. Hence the continuum modes $f_{nLi}(y)\psi_{nLi}(x) $ and their counterparts $f_{nRi}(y)\psi_{nRi}(x)$, which have analogous conditions omitted here for simplicity, constitute properly normalized solutions to the Dirac equation.

Thus there is a continuum of massive chiral 3+1-dimensional modes starting from zero mass. It is argued that provided $0 < \frac{\partial p}{\partial y} << 1$ the potential, $e^{-p(y)/6M^3} W$, will decay slowly to 0. This provides a wide barrier for the low energy, asymptotically free, continuum modes to tunnel through. Hence the corresponding wave functions will be heavily suppressed at the position of the brane: $y=0$. The same behavior is exhibited by the spectrum of Kaluza-Klein modes for the general linearlized fluctuations around the metric in the Randall-Sundrum delta function brane case \cite{Randall:1999vf}. We argue that it is possible for the cross section of any process involving interaction between the zero mode and light continuum modes to be imperceptibly low.

Because the $\psi_{n,L\setminus R, i}$ are not ${\rm SO}(10)$ gauge singlets, individual modes will propagate as constituent particles of gauge singlet states in the confining bulk. However as argued in section (\ref{sec:Localisingfermions}) this does not compromise the integrity of our choice of mode decomposition for $\Psi(x,y)$. Furthermore bound states comprised of low energy, massive, 3+1-dimensional chiral, continuum modes will propagate far from the brane, while the zero mode fermion will be trapped around the 3+1-dimensional topological defect. Hence the cross section for processes involving interactions between the zero mode and the low energy continuum modes will still be extremely low.

\subsection{Additional Symmetry Breaking}
\label{sec-aditionalsymetrybreaking}

The final components of our model are the ${\rm SU}(5)$-breaking and
electroweak breaking Higgs fields, $\zeta(x^{\mu},y)$ and $\eta(x^{\mu}, y)$ respectively. We introduce these fields now. There are many well documented ways of breaking ${\rm SU}(5) \times {\rm U}(1)_X$ down to the electroweak gauge group and beyond. We choose a specific scenario to illustrate the general process.

\noindent{The symmetry breaking pattern:}
 \begin{align}
 {\rm SO}(10) & \supset {\rm SU}(5) \times {\rm U}(1)_X \notag\\
 &\longrightarrow  {\rm SU}(3)_C \times {\rm SU}_W(2) \times {\rm U}_Y (1) \notag\\
 & \longrightarrow  {\rm SU}(3)_C \times {\rm U}(1)_Q
 \end{align}
 \noindent{can easily be achieved by using a pair of 16 dimensional representations for ${\rm SO}(10)$. The branching rules for this representation are \cite{Slansky:1981yr}:}
 \begin{eqnarray}
{\rm SO}(10) &\supset & {\rm SU}(5) \times {\rm U}(1)_X \notag \\
16 & \rightarrow & 10(-1) + 5^*(3) + 1(-5) \notag \\
{\rm SO}(10) &\supset& {\rm SU}(3)_C \times {\rm SU}(2)_W \times {\rm U}(1)_Y \times {\rm U}(1)_X \notag \\
16   & \rightarrow & (3,2)(1)(-1) + (3^*,1)\left(-4\right)(-1) +
   (1,1)(6)(-1) +
    (3^*,1)(2)(3) + (1,2)(-3)(3) +
   (1,1)(0)(-5).
\end{eqnarray}
\noindent{We would like the component of $\zeta(x^{\mu},y)$ which transforms like $(1,1)(0)(-5)$ under ${\rm SU}(3)_c \times {\rm SU}(2)_W \times {\rm U}(1)_Y \times {\rm U}(1)_X$ to condense inside the domain wall. The other 15 components of $\zeta(x^{\mu},y)$ should not condense. This will ensure that the gauge group is broken to the standard model on the brane.}

Similarly for the field $\eta (x^{\mu}, y)$, we would like the field component which is uncharged under the embedded $ {\rm U}(1)_Q$ belonging to the doublet $(1,2)(-3)(3)$ to condense inside the domain wall and all other components to not condense. This will implement electroweak symmetry breaking on the brane.
 The most general 4th order ${\rm SO}(10)$ invariant potential felt by these Higgs fields is
 \begin{align}
V_{\zeta, \eta} = \Sigma_{i,j} \{\lambda_{H1,h_i, h_j} h_i^{\dagger} h_j + \lambda_{H 2,h_i,h_j} \left(h_i^{\dagger}h_j\right)^2  + \lambda_{H 3} \left(h_i^{\dagger}h_i\right)\left(h_j^{\dagger}h_j\right) + \lambda_{H 4, h_i,h_j}h_i^{\dagger}h_j \phi^2
 \notag \\ + \lambda_{H 5, h_i, h_j} h_i^{\dagger}h_j {\rm Tr} {\cal X}^2
+ \lambda_{H 6, h_i, h_j} h_i^{\dagger} (\tau^a\tau^b {\cal X}_{ab})^2 h_j +\lambda_{H 7, h_i, h_j}h_i^{\dagger} \tau^a\tau^b {\cal X}_{ab} \phi h_j \},
  \label{eq-additionalhiggspotential}
 \end{align}
 \noindent{where the indices $i,j$ run over the set $\{1,2\}$ with $h_1 = \zeta(x^{\mu}, y)$ and $h_2 = \eta(x^{\mu}, y)$.}

 We factor the $k{\rm th}$ entry in the 16-dimensional column vector for each Higgs field into the product of an extra dimensional profile function $g_{h_i,k,n}(y)$ and a complete set of solutions to the 3+1-dimensional massive Klein-Gordon equations, $\theta_{h_i,k,n}(x^{\mu})$. That is,
 \begin{eqnarray}
 \begin{array}{l c r}
 [h_i]_k = \sum\hspace{-0.3 cm}\int_n g_{h_i,k,n}(y)\theta_{h_i,k,n}(x^{\mu}), & ~~\textrm{where}~~ & \partial_{\mu}\partial^{\mu} \theta_{h_i,k,n} + m^2_{n_{h_i,k}}\theta_{h_i,k,n} = 0.
 \end{array}
 \end{eqnarray}
\noindent{Thus if we let $\frac{\partial V_{\zeta, \eta}}{\partial [h^{\dagger}_i]_k} = U_{h_i,k,h_j,p} [h_j]_p + {\rm O}(h^2_i)$, where the indices $k$ and $p$ are being used to label the 16 components of each Higgs field, while $i$ and $j$ still run over the set $\{1,2\}$ with the same definition as before for $h_1 (x^{\mu},y)$ and $h_2 (x^{\mu}, y)$, then we must find the solutions to the Euler-Lagrange equations:}
\begin{equation}
-\frac{\partial^2 g_{h_i,k,n}(y)}{\partial y^2}\theta_{h_i,k,n}(x) + \frac{p'}{3M^3} g_{h_i,k,n}(y)\theta_{h_i,k,n}(x) + U_{h_i,k,h_j,p} g_{h_j,p,n}(y)\theta_{h_j, p,n}(x) = m^2_{n_{h_i,k}} e^{p(y)/6M^3}g_{h_i,k,n}(y)\theta_{h_i,k,n}(x).
\label{eq-eulerlagrangeequationsadditionalhiggs}
\end{equation}
\noindent{To get a qualitative idea of the behavior of solutions we consider the simplest case of equation (\ref{eq-eulerlagrangeequationsadditionalhiggs}). That is we choose the coupling constants in equation (\ref{eq-additionalhiggspotential}) so that $U_{h_i,k,h_j,p} = 0 ~\textrm{when}~ k \neq p \textrm{ or } h_i \neq h_j$. We would like the appropriate symmetry breaking components of $\eta (x^{\mu}, y)$ and $\zeta(x^{\mu},y)$ to have a Kaluza-Klein mode with tachyonic mass. This indicates an instability in the background solution $h_i = 0$. There is sufficient parameter freedom for this to be possible while keeping $\frac{p'}{3M^3} + U_{h_i,k,h_i,k} - m^2_{n_{h_i,k}} e^{p(y)/6M^3}$ as a potential well centered about the coordinate of the domain wall brane. This is necessary to ensure that the condensed component of each Higgs field is localised to the domain wall. For the remaining 15 components of each Higgs field it is only necessary to ensure that one of these two conditions is voided so that the field does not condense inside the domain wall. A thorough analysis would then require us to go back and solve for both the flat space domain wall configuration and the Einstein-Klein-Gordon equations consistently with this new background to ensure the backreaction of having additional Higgs fields with tachyonic components does not destabilize the brane. This presents a considerable computational task and we do not attempt it here.}

The zero mode fermions can now acquire masses through coupling to the ${\rm SU}(5)$ and electroweak breaking Higgs fields. Coupling constants in the effective 3+1-dimensional theory arise from the putative 5-dimensional coupling constant multiplied by the overlap integral of the bulk profile functions. This follows from integrating out the y-dependence of the terms in the 5-dimensional action. Hence fermions with different bulk profile functions will have different tree level masses. Since the fermion profile functions are split along the bulk according to their $U(1)_X$ charge the normal ${\rm SO}(10)$ tree level mass relations are reduced to the less phenomenologically infringing ${\rm SU}(5)$ mass relations.

\section{${\rm SO}(10) \rightarrow {\rm SU}(3)_C \times {\rm SU}(2)_L \times
{\rm SU}(2)_R \times {\rm U}(1)_{B -L}$ Model Modifications}
\label{sec:SO(!0)breakstoSU(3)timesSU(2)LtimeSU(2)RtimesU(1)}

We would like to engineer a Higgs potential which breaks ${\rm SO}(10)
\rightarrow {\rm SU}(3)_C \times {\rm SU}(2)_R \times {\rm SU}(2)_L
\times {\rm U}(1)_{B-L}$ directly with the background ${\rm SO}(10)$ adjoint
Higgs. Unfortunately, for our fourth order potential, any Higgs
solution respecting this symmetry on the domain wall is
perturbatively unstable against further condensation. This point is considered in more detail in the Appendix (\ref{sec-apendix}). However it is possible to break ${\rm SO}(10)$ down to the left right symmetric model using a sixth order potential:
\begin{align}
 V_{{\cal X}, \phi} = -\frac{\mu^2}{2} {\rm Tr}{\cal X}^2 +
\frac{\lambda_1}{4}({\rm Tr}{\cal X}^2)^2  + \frac{\lambda_2}{4}{\rm Tr}{\cal X}^4 +
\frac{\lambda_3}{4}{\rm Tr}{\cal X}^6  + \frac{\lambda_4}{4}{\rm Tr}{\cal X}^2 {\rm Tr}{\cal X}^4 +
\frac{\lambda_5}{4}({\rm Tr}{\cal X}^2)^3 \notag \\ +\frac{\lambda}{4}(\phi^2
- \vee^2)^2(\phi^2 - \xi^2) +
\frac{\kappa}{4}\phi^2{\rm Tr}{\cal X}^2 +
\frac{\lambda_6}{4}({\rm Tr}{\cal X}^2)^2\phi^2 +
\frac{\lambda_7}{4}{\rm Tr}{\cal X}^4\phi^2 +
\frac{\beta}{4}\phi^4{\rm Tr}{\cal X}^2.
\label{eq-sixthorderpotential}
\end{align}
This introduces a plethora of dimension-full coupling constants. However, since we are unable to renormalize 4+1-dimensional Yang-Mills gauge theories, we have assumed there is an inherent cutoff scale beyond which our model is no longer effective, so we are at liberty to add sixth-order terms.

The Euler-Lagrange equations still exhibit flat space solutions of the form
\begin{equation}
 \phi(y) = \vee \tanh[my], \qquad {\cal X}_1 =
A{\rm sech}[my],
\label{eq-sithordersolutionseulerlagrane}
\end{equation}
\noindent{with $m^2 = -2\mu^2 + \kappa \vee^2 + \beta\vee^4$ and $A^2 = \frac{8(2 \mu^2 - \kappa \vee^2 - \beta \vee^4) - 4 \lambda\vee^2\xi^2 + 4\lambda \vee^4}{\kappa + 2 \vee^2\beta}$. We have relabeled the ${\rm SO}(10)$ generators so that $T_1$ now refers to $T_1 = \frac{1}{\sqrt{12}}
{\rm diag}(\sigma_2,\sigma_2,\sigma_2, 0,0)$.  However the
conditions are now a lot more convoluted with the solution
contingent on:}
 \begin{eqnarray}
(\lambda_3 + 2\lambda_4 + 36\lambda_5)A^4
 - (48\lambda_6\vee^2 + 8\lambda_7\vee^2)A^2 +
48\beta\vee^4 & = & 0\notag,\\
 12m^2 +\left(6\lambda_1 + \lambda_2\right)A^2 - 6\kappa \vee^2 + 6\lambda_6 \vee^2 A^2 + \lambda_7 \vee^2 A^2 -12\beta\vee^4  & = & 0, \notag\\
 72\lambda\vee^4 + \left(6\lambda_6+\lambda_7\right)A^4
- 24\vee^2\beta A^2 & = & 0.
\end{eqnarray}
A general perturbative linear analysis must now be done
numerically, as is explained in the Appendix (\ref{sec-apendix}). However we note that for the parameter regime
 \begin{eqnarray}
 2\lambda_1 A^2 + \lambda_2 A^2 - 2\kappa \vee^2 +
2\lambda_6 A^2\vee^2 + \lambda_7 A^2\vee^2 - 4\beta\vee^4  & > & 0, \notag \\
( 5\lambda_3 A^4 + 14\lambda_4 A^4 + 36\lambda_5 A^4 - 48\lambda_6\vee^2A^2 -
24 \lambda_7 \vee^2A^2 +
  48 \beta\vee^4 ) & > & 0,
\label{eq-sixthorderpotentialpositivityconditions}
\end{eqnarray}
\noindent{it is possible to guarantee that all perturbations about (\ref{eq-sithordersolutionseulerlagrane}) are oscillatory and ${\rm SO}(10)$ will break stably to ${\rm SU}(3)_C \times {\rm SU}(2)_L \times {\rm SU}(2)_R \times {\rm U}(1)_{B-L}$. We cannot predict the behavior of the perturbations outside this parameter regime. A general method for identifying the bifurcation points has yet to be determined. The fermions are localised by exactly the same
techniques as in the ${\rm SU}(5)\times {\rm U}(1)_X$ case.}
\def\epsfsize#1#2{0.8#2}
\begin{figure}[htbp]
\begin{center}
\epsfbox{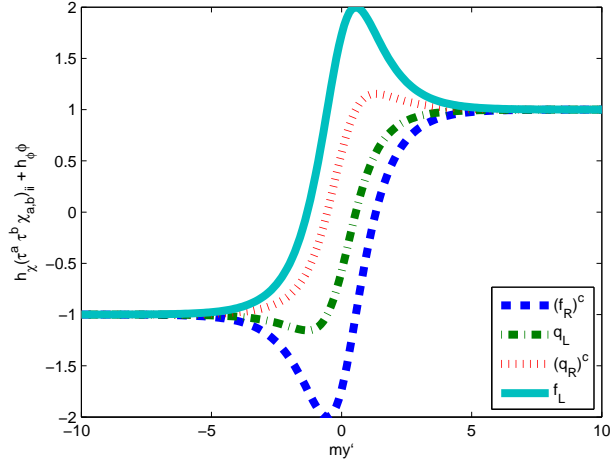}\\
 \caption{The graph displays the integrands in the exponent of equation (\ref{eq:fermionprofiles}) associated with extra dimensional profile functions for the left right symmetric model bulk fermions. The graph shown here corresponds to the parameter choices $h_{\phi} = h_{\cal X} =1$ and the flat space solution to the Euler-Lagrange equations ${\cal X}_1 = A{\rm sech}(my)$, $\phi = \vee{\rm tanh} (my)$ with $A = \vee = 1$. 3+1-dimensional left chiral fermions will be confined to the hyperplane associated with the zero of their integrand.}
\label{left-rightsymmetricfermions}
\end{center}
\end{figure}

The zero mode fermion profiles are now localised about the $x$-axis intercepts of Figure \ref{left-rightsymmetricfermions}.

We can also solve the Einstein-Klein-Gordon equations under these
circumstances. We find that our previous ansatz (\ref{eq-eienstienequationsolutions}), will satisfy the Einstein-Klein-Gordon equations for the sixth order potential provided we impose the same bulk cosmological constant fine tuning and we are in the slice through the parameter regime :
\begin{eqnarray}
-3\lambda\xi^2 - 12 \beta \vee^2 + 3 \lambda\vee^2 -6\kappa +
12\vee^2\lambda_6 + 2\lambda_7\vee^2 +
2\lambda_2 + 12\lambda_1 & = & 0, \notag\\
3M^3m^2 + \vee^2m^2 + 6M^3\mu^2 -3M^3\kappa\vee^2 - 3M^3\beta\vee^4 & = & 0,\notag\\
\lambda_3 + 6\lambda_4 + 36\lambda_5 - 12\lambda_6
- 2\lambda_7 + 3\beta & = & 0, \notag\\
6M^3 m^2 + \vee^2m^2 - 3M^3\lambda\vee^4 + 3M^3\lambda\vee^2\xi^2 +
3M^3 \kappa\vee^2 + 6M^3\beta\vee^4 & = & 0, \notag\\
9\lambda +12\lambda_6 +2\lambda_7 -12\beta & =
&0.
\end{eqnarray}
This symmetry breaking Higgs pattern selects a left right
symmetric model on the brane \cite{Senjanovic:1975rk}. It is subsequently possible to break our left right symmetric model down to the standard model by a technique similar to the one outlined in section (\ref{sec-aditionalsymetrybreaking}).

\section{Further Remarks}

There are a couple of salient points for 4+1-dimensional domain wall brane models, based on the Dvali-Shifman mechanism, which we have not yet considered. Firstly, since we are constructing an ${\rm SO}(10)$ grand unified theory we need the standard model coupling constants to unify at some high scale. Secondly, it was mentioned in section (\ref{sec:Dvali-Shifman gauge field localisation}) that we need a hierarchical ordering of the inherent physical scales for our model to work. Both these points were discussed in the DGV paper \cite{Davies:2007xr}. However in going from ${\rm SU}(5)$ to ${\rm SO}(10)$ gauge invariance in the bulk we have introduced a new intermediate symmetry breaking scale. For this reason we explicitly review both ideas in the context of our model.

The spectrum of Kaluza-Klein modes will affect standard model gauge coupling constant running. In particular Kaluza-Klein modes for 4+1-dimensional fields carrying different ${\rm U}(1)_X$ charges which belong to the same ${\rm SO}(10)$ multiplet in the unified ${\rm SO}(10)$ gauge theory generally have different masses, i.e. the Kaluza-Klein modes originate in the ${\rm SO}(10)$ gauge theory as split ${\rm SO}(10)$ multiplets. Because the Kaluza-Klein modes form split ${\rm SO}(10)$ multiplets, and at specific energy scales only certain ${\rm U}(1)_X$ components of the split multiplet will be able to contribute to the beta functions, the comparative running of the standard model coupling constants will change. Although coupling constant unification is ruled out for 3+1-dimensional non-supersymmetric ${\rm SO}(10)$ GUTs, it is still possible for our model. The full calculation would require us to do a phenomenological parameter fitting: this will fix the higher mass Kaluza-Klein modes and therefore determine the coupling constant running. We have not undertaken this particular task, however we find it tantalizing that the coupling constant may unify at some high scale \cite{RDavies2007}.

 To maintain the internal consistency of our model, we require a specific ranking of the magnitudes of the inherent physical scales. For simplicity we will look specifically at the ${\rm SO}(10) \rightarrow {\rm SU}(5)$ model. The concepts do not change for the ${\rm SO}(10) \rightarrow {\rm SU}(3)_C \times {\rm SU}(2)_L \times {\rm SU}(2)_R \times {\rm U}(1)_{B -L}$ model, hence everything we say here is relevant to this model as well.

We need to look carefully at 5 specific physical scales: the ultraviolet cut off $\Lambda_{\rm UV}$, the ${\rm SO}(10)$ and ${\rm SU}(5)$ breaking scales on the brane $\Lambda_{{\rm S0}(10)} \sim [{\cal X}(my =0)]^{2/3}$ and $\Lambda_{{\rm SU}(5)} \sim \langle \zeta_{(1,1)(0)(-5)} (my = 0) \rangle^{2/3}$ respectively, the ${\rm SO}(10)$ confinement scale in the bulk $\Lambda_{\rm conf}$ and the inverse wall width $\Lambda_{\rm DW} \equiv m$. These scales must be ranked according to:

\begin{equation}
\label{eq:heirachyofscales}
\Lambda_{\rm DW} < \Lambda_{\rm conf} < \Lambda_{{\rm SU}(5)} < \Lambda_{{\rm SO}(10)} < \Lambda_{\rm UV}.
\end{equation}

This hierarchy is necessary for the integrity of the Dvali-Shifman mechanism. Lattice gauge theory simulations \cite{Laine:2004ji} tell us that the width of the domain wall must be greater than the ${\rm SO}(10)$ glueball radius for the Dvali-Shifman mechanism to work. This translates into the relation: the inverse wall width is less than the bulk confinement scale.

The confinement scale in the bulk must be lower than the ${\rm SO}(10)$ and ${\rm SU}(5)$ symmetry breaking scales on the brane. Otherwise the non-perturbative bulk ${\rm SO}(10)$ confinement dynamics would dominate on the brane and the behavior of the $\zeta$ and ${\cal X}$ fields which set these scales would be dictated by their strong ${\rm SO}(10)$ interactions. In this case the classical dynamical equations we have solved for our background $\zeta$ and ${\cal X}$ configurations would not be appropriate and our theory would have an internal inconsistency.

Finally our model is an effective field theory which is assumed to be valid only up to the ultraviolet cut off. This establishes $\Lambda_{\rm UV}$ as the highest energy scale in the theory.

Amongst the above scales, $\Lambda_{\rm DW}$, $\Lambda_{{\rm SU}(5)}$, $\Lambda_{{\rm S0}(10)}$ and $\Lambda_{\rm UV}$ are all functions of the free parameters in the Lagrangian. We argue that there is sufficient parameter freedom to rank them according to (\ref{eq:heirachyofscales}). The bulk ${\rm SO}(10)$ confinement scale is putatively to be calculated from the value of the dimensionful gauge coupling constant $g$ for the bulk ${\rm SO}(10)$ gauge theory. Because this gauge theory is not renormalisable in 4+1-dimensions and we have introduced a UV cut off, $\Lambda_{\rm conf}$ will depend on both $\Lambda_{\rm UV}$ and $g$. If we follow the trend from section (\ref{sec:Dvali-Shifman gauge field localisation}) and assume that our model has a transition to a confining regime for values of the dimensionful gauge coupling constant $g$, greater than a critical value $g_c\left(\Lambda_{\rm UV}\right)$, then the bulk confinement scale will be set by $\Lambda_{\rm UV}$ and a $g > g_c\left(\Lambda_{\rm UV}\right)$. This calculation is beyond the scope of the present work.

\section{conclusion}
We have identified all the ingredients needed for a 4+1-dimensional field
theoretical model to produce an effective theory of 3+1-d fields localised to a domain wall brane with
${\rm SO}(10)$ gauge invariance in the bulk. We have presented two
models, whereby the symmetry on the brane is spontaneously broken
to give an ${\rm SU}(5)\times {\rm U}(1)_X$ unified model and a left
right symmetric model respectively. Our models also exhibit
localisation of the zero mode linearized fluctuation about the
metric. We reiterate that the signature points of our model
are:

\begin{itemize}
\item{All 4 spatial dimensions are infinite.}
\item{The brane is supplied by a dynamically generated domain
wall.}
\item{Standard Model fields are localised by their interactions
with the domain-wall-engineering Higgs fields.}
\item{All Standard Model fermions get unified in a single ${\rm SO}(10)$ representation in the bulk.
However ${\rm SO}(10)$ mass relations are conspicuously absent from
our model.}
\end{itemize}

Further investigation into the implementation of Dvali-Shifman
gauge field localisation arguments in 4+1-dimensions will be
necessary to establish a rigorous foundation for future models.



\medskip

{\bf Acknowledgments:}
We thank R. Davies, D. P. George and N. Pesor for useful discussions
and help with computational work. JET also thanks H. B. Thompson and Michal Feckan for help with constraining eignevalue spectrums for ODEs. This work was supported by the
Australian Research Council and the Puzey Bequest to the University of Melbourne.
%

\section{Appendix A}
\label{sec-apendix}
We are interested in the stability of a solution to the Euler-Lagrange equations following from equation (\ref{eq: fourthorderhigspotential}),
\begin{eqnarray}
0 & = &  \Box \phi + \frac{\kappa}{2}{\cal X}_{ij}{\cal X}_{ji}\phi + \lambda \phi \left(\phi^2 -\vee^2 \right), \notag\\
0 & = & \Box {\cal X}_{ab} - 2\mu^2 {\cal X}_{ab} + 2\lambda_1 {\cal X}_{ab} \left({\cal X}_{ij}{\cal X}_{ji}\right)+
2\lambda_2\left({\cal X}^3\right)_{ab} + \kappa \phi^2{\cal X}_{ab}.
\label{eq-SU(5)eulerlagrangeequations}
\end{eqnarray}
\noindent{We shall call this solution (${\cal X}_0(y), \phi_0(y)$). For the cases examined in this paper we shall only consider ${\cal X}_0(y)$ for which all the nonzero components are in the directions of the elements of the Cartan subalgebra for ${\rm SO}(10)$ which we explicitly describe by $C_1 = \frac{1}{2}\textrm{diag}(\sigma_2, 0 ,0 ,0,0), \dots C_5 = \frac{1}{2}\textrm{diag}(0,0,0,0,\sigma_2)$. To establish the stability of this solution we consider the dynamical evolution of a perturbation $(\delta\bar{{\cal X}}(x,y,t), \delta\phi(x,y,t))$ in a background formed by $({\cal X}_0(y),\phi_0(y))$. It is argued that if a perturbation is not allowed because of the symmetry of the background domain wall brane configuration or if a perturbation has oscillatory time dependence so that $\frac{||\delta\bar{{\cal X}}(x,y,t)||}{||{\cal X}_0(y)||}, \frac{||\delta\phi(y,t)||}{||\phi_0(y)||} << 1, ~ \forall t > 0 $, then the domain wall brane formed by (${\cal X}_0(y), \phi_0(y)$) is stable. The perturbation $\delta\bar{{\cal X}}(y,t)$ is allowed to vary independently in the direction of each of the 45 components of the adjoint representation for ${\rm SO}(10)$. We use the notation $\delta\bar{{\cal X}}(x,y,t) = \delta{\cal X}(x,y,t) + N(x,y,t)$, where $\delta{\cal X}(x,y,t)$ is a matrix containing the perturbations along the directions of the Cartan subalgebra generators for ${\rm SO}(10)$, and $N(x,y,t)$ contains all the perturbations in the subspace of the adjoint representation that is orthogonal to the Cartan subalgebra under the trace operator. Thus $ N(x,y,t)$ is a matrix with $2 \times 2$  blocks of zero matrices along the diagonal.

We will break this argument down by ruling out classes of perturbations. We will start by arguing that perturbations which break the 3-dimensional rotational invariance respected by the solution (${\cal X}_0(y), \phi_0(y)$) are not allowed and hence we will suppress the coordinate label x and write $\left({\cal X}(y,t), \phi(y,t)\right) = \left({\cal X}_0(y,t) + \delta{\cal X}(y,t) + N(y,t), \phi_0(y,t) + \delta \phi(y,t)\right)$. We then successively rule out all perturbations to ${\cal X}_0 (y)$ of the form $N(y,t)$. Finally we will show that all perturbations of the form $\left(\delta {\cal X}(y,t), \delta \phi (y,t)\right)$ are oscillatory in time and hence the background solution (${\cal X}_0(y), \phi_0(y)$) is stable under dynamical evolution.

Since our solution (${\cal X}_0(y), \phi_0(y)$) is invariant under 3-dimensional spatial rotations there is no preferred direction in the hyperplane orthogonal to the $y$-coordinate. We argue that because the background solution treats all spatial directions orthogonal to the bulk coordinate, $y$, as the same, it does not make sense to say a perturbation has formed along the direction of a specific vector in this orthogonal space. This is because all directions are relative to an arbitrary choice of reference coordinate system \cite{Pogosian:2000xv}. Hence we only consider perturbations of the form $\left(\delta {\cal X}(y,t) + N(y,t), \delta \phi (y,t)\right)$.

It is argued in \cite{Pogosian:2001fm} that if a perturbation does not change the energy density of our solution (${\cal X}_0(y), \phi_0(y)$) by introducing a term which is linear in the perturbative fields then we can have a stable solution where these perturbations are set to zero in the expansion for $\left({\cal X}(y,t), \phi(y,t) \right)$.

A term which is linear in the perturbative fields would be introduced through an expansion of the ${\rm SO}(10)$ Casimir invariants appearing in the potential. Hence we consider the form of ${\rm Tr}{\cal X}^2$ and ${\rm Tr}{\cal X}^4$ for ${\cal X}(y,t) = {\cal X}_0(y) + \delta\bar{{\cal X}}(y,t) = \{{\cal X}_0(y) +  \delta{\cal X}(y,t)\} + N(y,t)$ which we will shorten to ${\cal X} = \{{\cal X}_0 +  \delta{\cal X}\} + N$ for convenience. For ${\rm Tr}{\cal X}^2$ the expression is
\begin{equation}
{\rm Tr}\left[\{{\cal X}_0 +  \delta{\cal X}\} + N\right]^2 = {\rm Tr}\left[\{{\cal X}^2_0 + 2 {\cal X}_0  \delta{\cal X} +  (\delta{\cal X})^2\} + \{{\cal X}_0 +  \delta{\cal X}\}N +  N^2\right].
\end{equation}
Clearly this Casimir invariant has introduced a term which is linear in the perturbations along the 5-independent Cartan subalgebra direction $C_1, \dots, C_5$. Hence we must examine perturbations along the Cartan subalgebra directions in more detail to determine their stability. However for the perturbations along the direction of the other 40 adjoint Higgs field generators, orthogonality of the ${\rm SO}(10)$ generators under the matrix bilinear form, trace, means that the linear term disappears. In the above equation this is ${\rm Tr}\left[{\cal X}_0 N\right] = 0 $. A similar observation can be made for all the Casimir invariants for ${\rm SO}(10)$ because the only possible terms in a perturbative expansion which are first order in the perturbations, will be of the form ${\rm Tr}{\cal X}^{2k} \supset {\rm Tr}\left[{\cal X}^{2k-1}_0 (\delta{\cal X} + N)\right]$, where $k \in {\mathbb Z}^+$. The exponent, $n$, for a general Casimir invariants can always be written as $n = 2k$ because the antisymmetry of the ${\rm SO}(10)$ generators implies  ${\rm Tr}{\cal X}^{2k+1} =0$ . It is easy to check that for any solution  ${\cal X}_0$ written as fields distributed over $C_1, \dots, C_5$, ${\cal X}^{2k-1}_0$ will also contain nonzero components only in the direction of the  ${\rm SO}(10)$ Cartan subalgebra generators, $C_1, \dots, C_5$, and hence ${\rm Tr}\left[{\cal X}^{2k-1}_0 N\right] = 0$  \cite{Pogosian:2001fm}. This means that the ${\rm Tr}{\cal X}^4$ does not contribute any terms to the energy density that are linear in the perturbations $N(y,t)$. So a stable solution $\left({\cal X}(y,t), \phi(y,t) \right)$ can be constructed using only the Cartan subalgebra generators and we do not need to worry about linear perturbations along the directions of ${\rm SO}(10)$ generators which are orthogonal to the Cartan subalgebra $C_1, \dots, C_5$.

It is left to establish that the perturbations $\delta{\cal X}(y,t)$, along the Cartan subalgebra directions $C_1, \dots, C_5$, have oscilatory time dependence under dynamical evolution. For definiteness we choose $\delta{\cal X}(y,t)$ to be the $10 \times 10$ matrix:}
\begin{equation}
\delta{\cal X}(y,t) = \delta{\cal X}_{12}(y,t) C_1 + \delta{\cal X}_{34}(y,t) C_2 + \delta{\cal X}_{56}(y,t) C_3
+ \delta{\cal X}_{78}(y,t) C_4~ +
\delta{\cal X}_{910}(y,t) C_5.
\end{equation}

\noindent{Thus in shorthand our perturbed solution is:}
 \begin{equation}
 {\cal X}(y,t)  = {\cal X}_0(y) + \delta{\cal X}(y,t), \qquad  \phi(y,t)  = \phi_0(y) + \delta\phi(y,t).
 \end{equation}
 \noindent{We substitute this expansion into (\ref{eq-SU(5)eulerlagrangeequations}) and argue that initially time evolution will be dictated by terms of first order in the small quantities $\delta \phi (y,t)$ and $\delta {\cal X}(y,t)$. Discarding all higher order terms we are left with coupled linear homogeneous equations, for $\delta\phi(y,t)$ and the  {\it a,b}th entry of $\delta{\cal X}(y,t)$, of the form:}
\begin{eqnarray}
0  &=&  \Box \delta\phi  + \kappa\phi_0\left({\cal X}_0\right)_{ij}\left(\delta{\cal X}\right)_{ji} +
\left(\frac{\kappa}{2} \left({\cal X}_0\right)_{ij}\left({\cal X}_0\right)_{ji}+ \lambda\left(3\phi^2_0 -\vee^2\right) \right)\delta \phi,
 \notag\\
0  &=&  \Box \delta{\cal X}_{ab} +  4\lambda_1\left({\cal X}_0\right)_{ab}\left({\cal X}_0\right)_{ij}\left(\delta{\cal X}\right)_{ji} +
 \left( -2\mu^2 + 2\lambda_1 \left({\cal X}_0\right)_{ij}\left({\cal X}_0\right)_{ji} + \kappa\phi_0^2 \right) \delta{\cal X}_{ab} +
 2\lambda_2 \left(\delta {\cal X}\right)_{ai} \left({\cal X}_0\right)_{ij} \left({\cal X}_0\right)_{jb}\notag\\
 &&+ 2\lambda_2\left({\cal X}_0\right)_{ai}\left(\delta{\cal X}\right)_{ij}\left({\cal X}_0\right)_{jb} +  2\lambda_2\left({\cal X}_0\right)_{ai}\left({\cal X}_0\right)_{ij}\left(\delta {\cal X}\right)_{jb}  +
 \left(2\kappa\phi_0\left({\cal X}_0\right)_{ab}\right)\delta\phi.
 \label{eq-perturbativeequations}
\end{eqnarray}
These equations can be reduced to ordinary differential equations by using Fourier decomposition to factor out the time dependence of $\delta \phi(y,t)$ and $\delta{\cal X}(y,t)$. We consider the evolution of a specific normal mode of the coupled system $(e^{\omega t} \delta {\cal X}(y), e^{\omega t}\delta\phi (y))$.

Each mode must independently satisfy the boundary conditions: $\delta {\cal X}(y), \delta \phi(y) \rightarrow 0$ as $y \rightarrow \pm \infty$ since a global change to the profile of either ${\cal X}(y)$ or $\phi(y)$ would require an infinite 3+1-dimensional energy density and could not be accomplished by perturbative effects alone.

If $ \omega^2 \le 0 $ for all coupled solutions $(e^{\omega t}\delta {\cal X}(y), e^{ \omega t} \delta \phi(y))$ of the above equations, then the perturbation will be oscillatory and the background solution $({\cal X}_0, \phi_0)$ will be perturbatively stable over short periods of dynamical evolution. Of course the solutions to  (\ref{eq-perturbativeequations}) depend on the parameters $\mu^2, \lambda_1, \lambda_2, \lambda, \kappa$. Thus we are looking for relations between the parameters which guarantee $\omega^2 \le 0$ for all normal modes.

Furthermore we are primarily interested in whether the symmetry breaking pattern ${\rm SO}(10) \rightarrow {\rm SU}(5) \times {\rm U}(1)_X$ is stable. Rather than tackling the full problem we will assess when our analytic solution (\ref{eq-SU(5)solutiontoeulerlagrangeequations}) is stable, since this will provide an example where ${\rm SO}(10)$ breaks to ${\rm SU}(5) \times {\rm U}(1)_X$ and stops there. Hence we perturb about $({\cal X}_0 (y), \phi_0 (y)) = (A {\rm sech} (my), \vee {\rm tanh} (my))$  and divide our problem into two scenarios: either all non-zero components of the superdiagonal of $\delta {\cal X}(y)$ are equal, or two such components, lets call them $-i \delta {\cal X}_{k,k+1}(y)$ and $-i \delta {\cal X}_{n, n+1}(y)$, are distinct.

In the first case, there is an unstable perturbation if and only if our solution (\ref{eq-SU(5)solutiontoeulerlagrangeequations}) is not the lowest energy background configuration for the Higgs fields which breaks ${\rm SO}(10)$ to ${\rm SU}(5) \times {\rm U}(1)_X$ and interpolates between the vacua $\phi = - \vee, {\cal X} = 0$ and $\phi = + \vee, {\cal X} = 0$.

We have numerically solved equation (\ref{eq-SU(5)eulerlagrangeequations}) using a relaxation method for a range of points $(\mu, \lambda_1, \lambda_2,\lambda, \kappa, \vee)$, in the free parametric space. We used a trial ansatz ${\cal X} = {\cal X}_{12} (y)/{\sqrt{5}}~ \left[C_1 + C_2 + C_3 + C_4 + C_5\right]$, $\phi = \phi(y)$ which automatically selects a ${\rm SU}(5) \times {\rm U}(1)_X$ symmetry on the domain wall. The results back up the claim: for a wide range of free parameters  the lowest energy solution to (\ref{eq-SU(5)eulerlagrangeequations}) which breaks ${\rm SO}(10) \rightarrow {\rm SU}(5) \times {\rm U}(1)_X$ and interpolates between the vacua $\phi = - \vee, {\cal X} = 0$ and $\phi = + \vee, {\cal X} = 0$ has the same qualitative form as (\ref{eq-SU(5)solutiontoeulerlagrangeequations}).

In the second scenario: there exists $ k, n \in \{1,...9\}$ such that $-i \delta {\cal X}_{k,k+1}(y)$ and $-i \delta {\cal X}_{n, n+1}(y)$, are distinct. Therefore we have a non-trivial differential equation for $\delta {\cal X}_{\epsilon} = \delta {\cal X}_{k,k+1} - \delta {\cal X}_{n, n+1}$. We seek conditions which guarantee $\omega^2 \le 0$ for all solutions to this equation:
\begin{equation}
-\frac{\partial^2 \delta {\cal X}_{\epsilon}}{\partial y^2} ~ - ~ \left({2m^2 - \frac {\lambda_2 A^2}{5}}\right) {\rm sech}^2 ( my ) ~\delta {\cal X}_{\epsilon} = \left(-\omega^2 -m^2\right)\delta {\cal X}_{\epsilon}.
\label{eq-schrodingerperturbationequation}
\end{equation}
\noindent{We change variables to $z=my$ so that we are working with a dimensionless co-ordinate system.
In this co-ordinate system we are able to rewrite equation (\ref{eq-schrodingerperturbationequation})
entirely in terms of a set of parameters with zero mass dimensions $\{ \lambda_1, \lambda_2, \lambda, \kappa, \vee^2, A^2, \omega^2\} \rightarrow \{\lambda_1 m, \lambda_2 m, \lambda m, \kappa m,  v^2 /m^3, A^2/m^3, \omega^2/m^2\}$. We note that because $m$ has to be real valued to ensure a solitonic background solution for $\phi (y)$ our condition for perturbative stability, in terms of the new parameter $\omega^2/m^2$, is $ \omega^2/m^2 \le 0$ . To make it easier to cross reference the results of this stability analysis with section (\ref{sec-domainwallconstructionandgaugefieldlocalisation}) we choose not to relabel these new dimensionless parameters. Instead of relabeling we point out that stability conditions written in terms of these new parameters will look exactly the same as conditions written in terms of the old dimensionful parameters if we set $m=1$. We hope it is clear that this is not a fine tuning condition; it is a shift of notation to a convention where all parameters and variables have zero mass dimension. Equation (\ref{eq-schrodingerperturbationequation}) becomes:
\begin{equation}
-\frac{\partial^2 \delta {\cal X}_{\epsilon}}{\partial z^2} ~ - ~ \left({2 - \frac {\lambda_2 A^2}{5}}\right) {\rm sech}^2 z ~\delta {\cal X}_{\epsilon} = \left(-\omega^2 - 1\right)\delta {\cal X}_{\epsilon}.
\label{eq-schrodingerperturbationequationm=1}
\end{equation}
\noindent{Let $U = \left({2 - \frac {\lambda_2 A^2}{5}}\right)$.  Equation (\ref{eq-schrodingerperturbationequationm=1}) is just a 1-dimensional time independent Schr\"{o}dinger equation and if we impose the condition $U > 0$ then negative eigenvalues $-\omega^2 < 0$ will correspond to bound states of the potential $-U  {\rm sech}^2 ( z ) + 1$. So in what follows we are simply determining the conditions necessary for this potential well to be sufficiently shallow so that it does not admit bound states.}

We use the following lemma from \cite{Yagasaki:1999}, to conclude that if $\lambda_2 > 0$ and $ \lambda_2 A^2 < 10 $ then there are no nontrivial solutions, to equation (\ref{eq-schrodingerperturbationequationm=1}), with $\omega^2 > 0$.
\begin{lemma}{\rm \cite{Yagasaki:1999}}\label{L9} Let $\kappa>0$, $\lambda>0$. The equation
\begin{equation}\label{eq:lemmaode}
\ddot v + (-\lambda + \kappa
\sech^2{t})v = 0
\end{equation}
has a bounded solution if and only if there
exists an integer $M$ such that
\begin{equation}
\begin{array}{rl}
\ds \lambda &= \tfrac14(\sqrt{4\kappa+1} - 4M -1)^2 \quad \text{for}
\quad
  0 \le M < \tfrac14(\sqrt{4\kappa+1} - 1) \\
\ds \textrm{or}\quad \lambda &= \tfrac14(\sqrt{4\kappa+1} - 4M -3)^2
\quad \textrm{for} \quad
  0 \le M < \tfrac14(\sqrt{4\kappa+1} - 3).
\end{array}\end{equation}\end{lemma}
Hence our analytic solutions (\ref{eq-SU(5)solutiontoeulerlagrangeequations}) to the Euler-Lagrange equations (\ref{eq-SU(5)eulerlagrangeequations}) will maintain a form which breaks ${\rm SO}(10) \rightarrow {\rm SU}(5) \times {\rm U}(1)_X$ under short term dynamical evolution provided $\lambda_2 > 0 $ and $ \lambda_2 A^2 < 10 $.

If $\lambda_2 < 0$ then $\omega^2 > 0$ and there will be a perturbative mode which grows with time. This indicates that our analytical solution (\ref{eq-SU(5)solutiontoeulerlagrangeequations}) will be unstable under dynamical evolution and it is not possible to stably break ${\rm SO}(10)$ to ${\rm SU}(5) \times {\rm U}(1)_X$ with this solution. This appears fortuitous as we would expect a cascade effect where the solution for ${\cal X}(y)$ evolved spontaneously into a form which would break ${\rm SO}(10)$ to a subgroup of  ${\rm SU}(5) \times {\rm U}(1)_X$ on the domain wall. However $\delta {\cal X}_{\epsilon}$ is given by the difference of any two arbitrary superdiagonal entries in $\delta {\cal X}(y)$. Hence if $\lambda_2 < 0$ then each nonzero component of $\delta {\cal X}$ will diverge relative to all the others. Thus what we will be left with is an ${\rm SO}(10)$ adjoint Higgs profile that breaks ${\rm SO}(10)$ down to ${\rm U}(1)^5$ on the domain wall brane.
Ultimately, mode mixing and other nonlinear effects arising from terms of higher order in $\delta {\cal X}$ and $\delta \phi$ will influence the long term evolution of the system. Assessing these effect is beyond the scope of our quantitative analysis. We fall back on our claim that numerical solutions to equation (\ref{eq-SU(5)eulerlagrangeequations}) for ${\cal X}$ and $\phi$ indicate that the analytic solutions (\ref{eq-SU(5)solutiontoeulerlagrangeequations}) are stable under long term dynamical evolution when $\lambda_2 > 0 $ and $ \lambda_2 A^2 < 10 $.

The second half of this paper deals with a sixth order potential for ${\cal X}$ and $\phi$, (\ref{eq-sixthorderpotential}), where the solution to the Euler-Lagrange equations creates a symmetry breaking pattern ${\rm SO}(10) \rightarrow {\rm SU}(3)_C \times {\rm SU}(2)_L \times {\rm SU}(2)_R \times {\rm U}(1)_{B-L}$.  We would like to repeat the above techniques with the analytic solution, (\ref{eq-sithordersolutionseulerlagrane}), to these Euler-Lagrange equations. Thereby we will have demonstrated a slice of the parameter space where ${\rm SO}(10)$ breaks to ${\rm SU}(3)_C \times {\rm SU}(2)_L \times {\rm SU}(2)_R \times {\rm U}(1)_{B-L}$ stably. While most of the results for perturbations of the form $\left(\delta \bar{\cal X}, \delta \phi\right) = \left(\delta{\cal X} + N, \delta \phi \right)$ discussed in the context of the 4th order case can be easily generalized to the 6th order potential, there will be a slight twist because we can no longer use lemma \ref{L9}.

One way of looking at lemma \ref{L9} is to make a co-ordinate substitution $s = {\rm tanh}z$ in equation (\ref{eq:lemmaode}). This will convert the equation (\ref{eq:lemmaode}) into a Riemann equation with regular singular points at $s = \pm 1$. The equation is an example of a hypergeometric differential equation for which the eigenspectrum is well known \cite{UvarovNikiforov}. The associated eigenvectors are orthogonal polynomials where the inner product is given by
\begin{equation}
\langle u,l\rangle  = \int^{1}_{-1} u(s) l(s)  ds =  \int^{\infty}_{-\infty} u(z) l(z) \sech^2 z dz.
\end{equation}
In the case of the 6th order equation if we make the same co-ordinate substitution $s = {\rm tanh}z$  in equation (\ref{eq-sixthorderperturbationequation}) we find that $s = \pm 1$ are no longer regular singular points and we are no longer able to easily construct a power series solution about these points. Hence we do not have polynomial solutions associated with eigenvalues which cause the power series to truncate at a finite degree and the previous ideas will not work under these circumstances.

We fix this hole in our analysis. Let $\delta {\cal X}_{\epsilon} = \delta {\cal X}_{j,j+1} - \delta {\cal X}_{n, n+1}$ be nontrivial then for $m=1$ we have:
\begin{eqnarray}
((-\omega^2) - 1) \delta {\cal X}_{\epsilon} &=& -\frac {\partial^2 \delta {\cal X}_{\epsilon}}{\partial z^2}  + \left(\lambda_1 A^2 +\frac{\lambda_2}{2}A^2 - \kappa \vee^2 +
\lambda_6 A^2\vee^2 + \frac{\lambda_7}{2}A^2\vee^2 - 2\beta\vee^4 \right) {\rm sech}^2 z \delta {\cal X}_{\epsilon} + \notag \\
 &&\left( \frac{5\lambda_3}{48} A^4 + \frac{7 \lambda_4}{24}A^4 + \frac{3\lambda_5}{4}A^4 - \lambda_6\vee^2A^2 - \frac{\lambda_7}{2} \vee^2A^2 +
  \beta\vee^4 \right){\rm sech}^4 z \delta {\cal X}_{\epsilon}.
  \label{eq-sixthorderperturbationequation}
\end{eqnarray}
This is just a 1-dimensional time independent Schr\"{o}dinger equation with potential
\begin{equation}
U = 1+ \left(\lambda_1 A^2 +\frac{\lambda_2}{2}A^2 - \kappa \vee^2 +
\lambda_6 A^2\vee^2 + \frac{\lambda_7}{2}A^2\vee^2 - 2\beta\vee^4 ) {\rm sech}^2 z +
 ( \frac{5\lambda_3}{48} A^4 + \frac{7 \lambda_4}{24}A^4 + \frac{3\lambda_5}{4}A^4 - \lambda_6\vee^2A^2 - \frac{\lambda_7}{2} \vee^2A^2 +
  \beta\vee^4 \right){\rm sech}^4 z.
\end{equation}
\noindent{If we impose the conditions featured in (\ref{eq-sixthorderpotentialpositivityconditions}) then the potential will be positive definite and the only admissible solutions will be bound states with positive energy, hence $  \omega^2 \le 0$ for all modes, under these conditions.}

\end{document}